\documentclass[sigplan,screen]{acmart}

\AtBeginDocument{%
  \providecommand\BibTeX{{%
    \normalfont B\kern-0.5em{\scshape i\kern-0.25em b}\kern-0.8em\TeX}}}

\acmPrice{15.00}
\acmISBN{978-1-4503-XXXX-X/18/06}

\usepackage{threeparttable}
\usepackage{algorithmic}
\usepackage{algorithm}
\usepackage{amsmath}
\usepackage{multirow}
\usepackage{tikz}

\copyrightyear{2024}
\acmYear{2024}
\setcopyright{acmlicensed}\acmConference[ICCAD '24]{IEEE/ACM International Conference on Computer-Aided Design}{October 27--31, 2024}{New York, NY, USA}
\acmBooktitle{IEEE/ACM International Conference on Computer-Aided Design (ICCAD '24), October 27--31, 2024, New York, NY, USA}

\begin{document}

\title{
AnalogGym: An Open and Practical Testing Suite for Analog Circuit Synthesis
}
\subtitle{\emph{(Invited)}}

\author{Jintao Li}
\affiliation{
\institution{University of Electronic Science and Technology of China}
\country{}
}
\email{j.t.li@i4ai.org}

\author{Haochang Zhi}
\affiliation{%
  \institution{Southeast University}
  \country{}
}
\email{hc.zhi1999@gmail.com}

\author{Ruiyu Lyu}
\affiliation{%
  \institution{Fudan University}
  \country{}
}
\email{rylv22@m.fudan.edu.cn}

\author{Zhenxin Chen}
\affiliation{%
  \institution{Guangzhou University}
  \country{}
}
\email{zhenxinchen@gzhu.edu.cn}

\author{Wangzhen Li}
\affiliation{%
  \institution{Fudan University}
  \country{}
}
\email{wangzhenli21@m.fudan.edu.cn}

 \author{Zhaori Bi\textsuperscript{\textdagger}}
 \affiliation{
   \institution{Fudan University}
   \country{}
   }
   \email{zhaori_bi@fudan.edu.cn}
   
 \author{Keren Zhu\textsuperscript{\textdagger}}
 \affiliation{
   \institution{Fudan University}
   \country{}
   }
   \email{kerenzhu@ieee.org}

\author{Yanhan Zeng}
\affiliation{%
  \institution{Guangzhou University}
  \country{}
}
\email{yanhanzeng@gzhu.edu.cn}

\author{Weiwei Shan}
\affiliation{%
  \institution{National Center of Technology Innovation for EDA}
  \country{}
  }
\email{wwshan@seu.edu.cn}

 \author{Changhao Yan}
 \affiliation{
   \institution{Fudan University}
   \country{}
   }
   \email{yanch@fudan.edu.cn}

 \author{Fan Yang}
 \affiliation{
   \institution{Fudan University}
   \country{}
   }
   \email{yangfan@fudan.edu.cn}

\author{Yun Li\textsuperscript{\textdagger}}
\affiliation{
   \institution{University of Electronic Science and Technology of China}
   \country{}
   }
\email{Yun.li@ieee.org}

 \author{Xuan Zeng\textsuperscript{\textdagger}}
 \affiliation{
   \institution{Fudan University}
   \country{}
   }
   \email{xzeng@fudan.edu.cn}

\thanks{\textsuperscript{\textdagger}Corresponding authors.}
\renewcommand{\shortauthors}{Jintao Li et al.}


\begin{abstract}
Recent advances in machine learning (ML) for automating analog circuit synthesis have been significant, yet challenges remain. 
A critical gap is the lack of a standardized evaluation framework, compounded by various process design kits (PDKs), simulation tools, and a limited variety of circuit topologies. 
These factors hinder direct comparisons and the validation of algorithms. 
To address these shortcomings, we introduced AnalogGym, an open-source testing suite designed to provide fair and comprehensive evaluations. 
AnalogGym includes 30 circuit topologies in five categories: sensing front ends, voltage references, low dropout regulators, amplifiers, and phase-locked loops. 
It supports several technology nodes for academic and commercial applications and is compatible with commercial simulators such as Cadence Spectre, Synopsys HSPICE, and the open-source simulator Ngspice. 
AnalogGym standardizes the assessment of ML algorithms in analog circuit synthesis and promotes reproducibility with its open datasets and detailed benchmark specifications. 
AnalogGym's user-friendly design allows researchers to easily adapt it for robust, transparent comparisons of state-of-the-art methods, exposing them to real-world industrial design challenges, and enhancing the practical relevance of their work. Additionally, we have conducted a comprehensive comparison study of analog sizing methods on AnalogGym, highlighting the capabilities and advantages of different approaches.
AnalogGym is available in the GitHub repository\footnote{\url{https://github.com/CODA-Team/AnalogGym}}.
The documentations are also available at \footnote{\url{http://coda-team.github.io/AnalogGym/}}.
\end{abstract}


\maketitle
\section{Introduction} \label{sec:intro}
\begin{figure}
    \centering
    \includegraphics[width=1\linewidth]{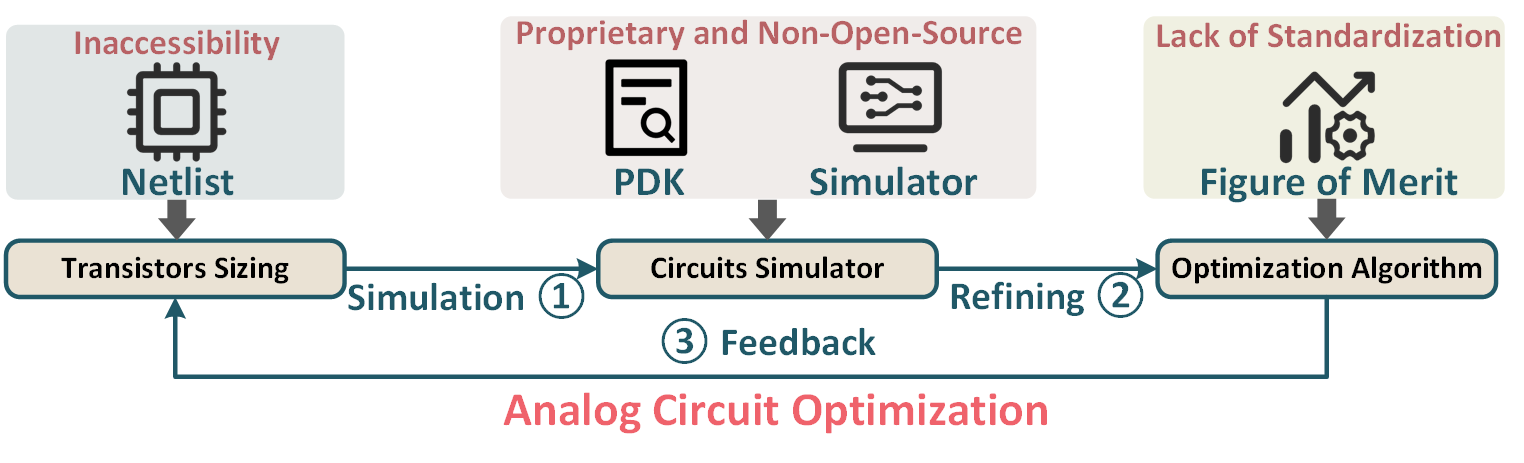}
    \vspace{-2em}
    \caption{Key challenges in analog circuit optimization}
    \vspace{-2em}
    \label{fig:Key Challenges}
\end{figure}

Automating the synthesis of analog integrated circuit (IC) through machine learning (ML) has shown significant promise, with various algorithms, such as Bayesian optimization (BO) \cite{2018TCASI:lyu, 2022TCAD:BO1, 2023ASPDAC:BO2,2024TCAD:Gu_BO,2024DATE:Gu_BO,2023DAC:cVTS_BO, 2021DAC_Touloupas}, reinforcement learning (RL) \cite{2020DAC:Wang, 2023ICCAD:GCN}, and evolutionary algorithms (EA) \cite{2021TCAD:ESSAB, 2024TCASI:Li, 2023ICCAD:Li, 2023MWSCAS_yanhan}, being applied to this challenging problem. 

Many research groups across the world have made varied contributions to the field.
The KU Leuven research group has significantly contributed through innovations in symbolic simulation, simulated annealing, and hierarchical system-level design synthesis \cite{1990ISSC_Gielen, 2003TCAD_Gielen, 2007IEEE_Gielen, 2000IEEE_Gielen}. 
The Fudan group has made notable advances in circuit optimization using Bayesian optimization algorithms, enhancing circuit sizing, yield optimization under the process, voltage, and temperature (PVT) variations, and high-dimensional performance modeling \cite{2024ASPDAC_zeng,2022ISCAS_zeng,2022MLCAD_zeng,2023TCAD_zeng,2022TCAD_zeng,2022TCAD_zeng1,2022TMTT_zeng}. 
Graeb et al. have developed important methods in structural modeling and topology recognition \cite{2021_Graeb2,2020ICCAD_Graeb1,2022ICCAD_Graeb3,2023PAINE_Graeb4}. 
Liu et al. at University of Glasgow have applied machine learning to global optimization, improving circuit sizing and microwave filter design \cite{2022TCAD_boliu1,2023TMTT_boliu}.
Ye et al. at Tsinghua University work on Python tools and DC-modeling has streamlined circuit optimization and topology prediction \cite{2023ISEDA_zuochang1,2023ISEDA_zuochang2,2023ISQED_zuochang3}. 
The UT group has also advanced design automation by utilizing deep learning for more efficient transistor sizing and topology selection \cite{2021DAC_David3,2021DATE_David2,2023ASPDAC_David1,2024DATE_David4, 2021DAC_zhu, 2019ICCAD_zhu, 20222ASPDAC_zhu}. 
These advancements highlight the potential of ML for assistant analog circuit design \cite{2023ICTA_ding, 2023ICAC_Li, 2019ASICON:li, 2020TCAD_zhaojun}, yet considerable challenges remain \cite{2024ASPDAC:lyu, 2024ASPDAC:xu, 2024ASPDAC:pan, 2023ISPD:pan, 2015ISPD_Rutenbar}.
The diversity of these methods reflects the vibrancy of the field.
However, it also reflects the lack of an open and accessible evaluation platform for comparing different methods.
The effort to build such a testing suite faces challenges related to accessibility and comparability as shown in Figure \ref{fig:Key Challenges}.

\textbf{Accessibility:} One of the primary challenges in analog circuit optimization is the lack of open-source process design kits (PDKs) and simulators. 
Proprietary PDKs with unique process parameters, transistor models, and design rules introduce inconsistencies that significantly impact performance metrics such as power consumption, speed, and area. These variations make it difficult to ensure that optimized designs are genuinely comparable. 
Without open-source PDKs and simulators, replicating results and verifying the effectiveness of different optimization approaches becomes increasingly challenging.

\textbf{Comparability:} 
Analog circuit specifications are complex and highly application-dependent, often involving dozens of performance metrics. 
For instance, when developing algorithms to optimize an amplifier (AMP), the lack of a comprehensive baseline might lead to a focus on metrics like power consumption and bandwidth, potentially causing the neglect of other critical performance factors such as transient response, which could undermine the effectiveness of the optimization.
Additionally, limited access to netlists undermines the accuracy and validity of comparisons across different optimization algorithms. Even when optimizing the same topology, inconsistencies in netlist details, such as whether complete biasing circuits are included, can lead to discrepancies in power.
A practical testing suite requires self-contained and comprehensive performance evaluation metrics for test circuits.

\begin{figure}
    \centering
    \includegraphics[width=1\linewidth]{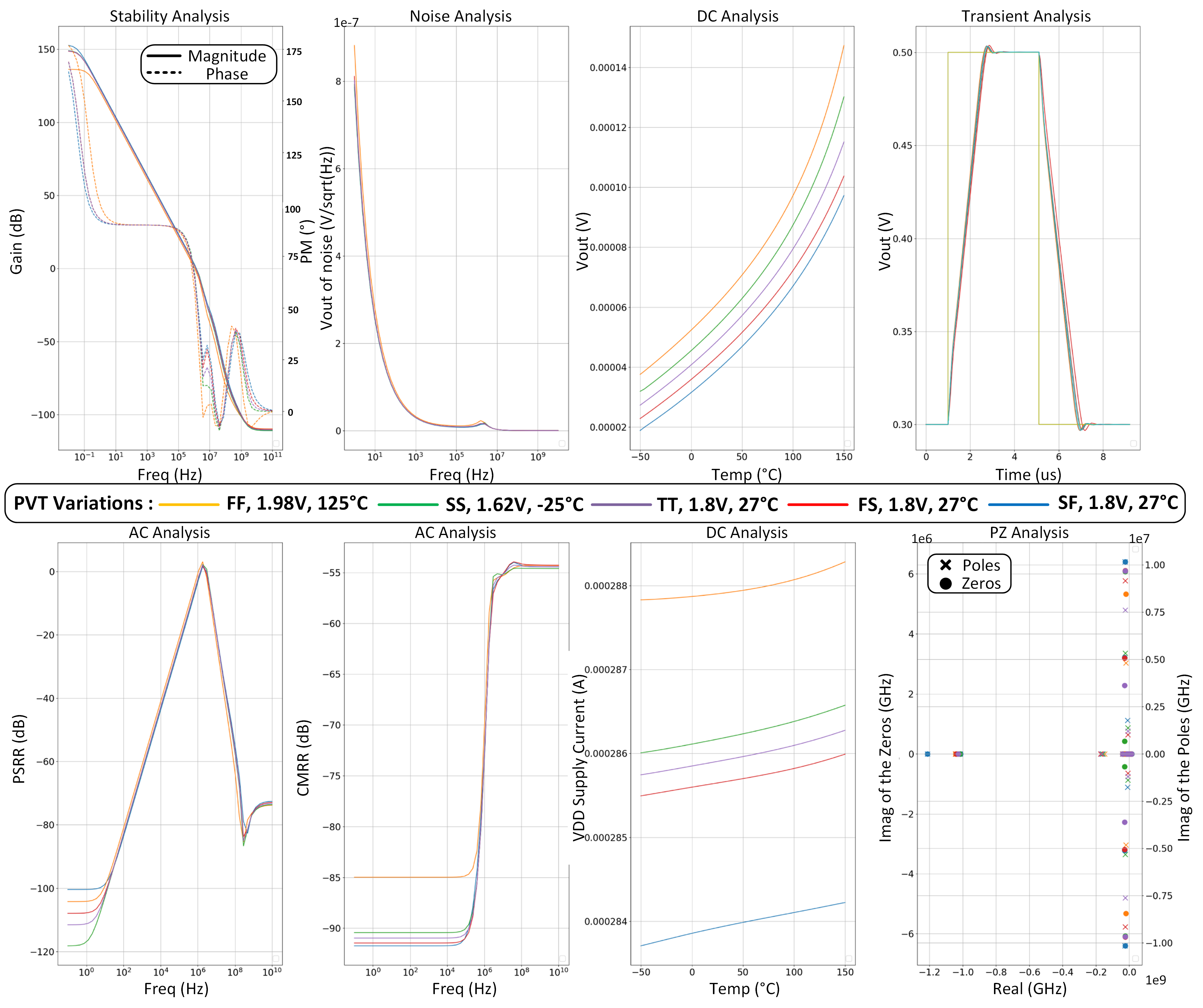}
    \caption{Comprehensive simulation results for amplifiers: DC, AC, PZ, and transient analyses}
    \vspace{-2em}
    \label{fig:Comprehensive_Simulation}
\end{figure}

To illustrate, consider the optimization of an AMP, as illustrated in Figure \ref{fig:Comprehensive_Simulation}. A comprehensive set of specifications must cover large-signal analysis, small-signal analysis, and PVT variations. To fully evaluate an amplifier's performance, simulations must include direct current (DC), alternating current (AC), pole-zero (PZ), and transient analyses. Without these comprehensive evaluations, critical aspects of the circuit's performance might be missed, leading to impractical designs.

Rapid advancement in fields such as computer vision can be largely attributed to the establishment of comprehensive benchmarks \cite{2001_cv, 2014_cv, 2020_cv, 2009:ImageNet}. These benchmarks provide a standardized way to evaluate and compare the performance of different algorithms, fostering innovation and collaboration. For example, ImageNet \cite{2009:ImageNet} has significantly advanced the field of computer vision by offering a large labeled dataset and clear evaluation metrics. This has enabled researchers to consistently measure improvements and identify state-of-the-art techniques, accelerating progress and ensuring reproducibility.

Drawing inspiration from the success of benchmarks like ImageNet and Gym \cite{2016_gym, 2024_gym}, it is evident that establishing a unified evaluation framework is crucial for analog circuit optimization. Such benchmarks should include open-source circuit netlists, testbenchs, standardized evaluation criteria, PDKs, and simulators. This would drive progress by enabling consistent and comparable assessments of different optimization approaches, providing a platform for researchers to test and validate their algorithms, fostering collaboration, and accelerating advancements in analog circuit design.

Therefore, we propose AnalogGym, an open and practical testing suite that provides a comprehensive and standardized framework for evaluating ML algorithms in analog circuit synthesis. AnalogGym encompasses 30 circuit topologies in five key categories: sensing front ends, voltage references, AMPs, low dropout regulators (LDOs), and phase-locked loops (PLLs). 
Among these, the LDOs and AMPs support the open-source Ngspice simulator and PDKs, allowing for greater accessibility and reproducibility. 

\begin{table}
\caption{Details of AnalogGym Components.}
\label{tab:Table of benchmark}
\renewcommand{\arraystretch}{1.1}
\resizebox{\linewidth}{!}{
\begin{tabular}{c|c|c|c}
\hline
Circuit Category  & Number of Topologies & Simulator & Open-source PDK \\ \hline
Amplifier              & 16 & Spectre/Ngspice & $\checkmark$ \\ \hline
Low dropout regulator  & 6  & Spectre/Ngspice & $\checkmark$ \\ \hline
Sensing front end      & 6  & Spectre         & $\times$ \\ \hline
Voltage reference      & 3  & Hspice          & $\times$ \\ \hline
Phase-locked loop      & 1  & Hspice          & $\times$ \\ \hline
\end{tabular}
}
\vspace{-2em}
\end{table}


The rest of the paper is organized as follows.
Section~\ref{sec:benchmark_descri} gives descriptions of AnalogGym.
Section~\ref{sec:Problem} introduces the problem formulation.
Section~\ref{sec:exp} evaluates several existing methods on AnalogGym.
Section~\ref{sec:discussion} discusses the paper and shows our perspectives.
In the end, Section~\ref{sec:conclusion} concludes the paper.

\section{Testbench Description} \label{sec:benchmark_descri}
We developed the testbench (TB) based on the open-source Ngspice simulator for the AMP and LDO circuits, incorporating the sky130 PDK\cite{s130_pdk_2024}. Voltage references and temperature sensing front ends do not support open-source PDKs due to the need for precise subthreshold modeling, which these PDKs often lack, leading to discrepancies in performance metrics. Similarly, PLLs require detailed analog and noise modeling that open-source PDKs typically do not provide, resulting in unreliable simulations of phase noise and jitter.

To streamline complexity and establish an open-source standard within the academic community, we have encapsulated the netlist, allowing for straightforward replacements within the TB. 
Note that the provided netlists include complete biasing circuits, which may result in higher overall power consumption.
The TB is provided in an easily interpretable textual format. 

A sample text file of TB on the NMCNR implemented with the sky130 PDK is shown in Table \ref{tab:Table of testbench}. The first line includes the encapsulated netlist, allowing for the selection of different experimental circuit netlists by modifying this line. The second, third, and fourth lines correspond to P, V, and T combinations. 
Adjusting the parameters in these lines allows selecting different PVT corners, as detailed in Table \ref{tab:Table of Physical Constraints}. 
The fifth line corresponds to the load capacitance of the circuit. Please note that the load capacitance for each circuit should match the load capacitance values in Table \ref{tab:baseline}.
The sixth line defines the parameters that require optimization within the circuit, including MOSFET size, capacitors, resistors, and input bias current.
It is important to note that in the sky130 PDK, MOSFET lengths and widths are specified without units and are understood to be in micrometers.

For MOSFETs, the naming convention follows a structured format using underscores to separate different attributes. For instance, in the name MOSFET\_23\_1\_L\_gm3\_NMOS:
\begin{enumerate}
    \item `MOSFET' indicates the component type.
    \item `23' is a randomly assigned index for identification purposes.
    \item `1' denotes the number of parallel transistors in the netlist.
    \item `L' specifies the parameter being described, which can be W (Width), L (Length), or M (Multiplicity).
    \item `gm3' describes the function of the transistor. The functions include: `gm' for transistors that convert the gate voltage to drain current, typically indicating the stage number in a multi-stage amplifier. For example, `gm3' denotes a transistor in the third stage responsible for this conversion. 
    `BIASCM' for bias current mirrors. 
    `LOAD' for load transistors. 
    \item `NMOS' indicates the transistor type, distinguishing between NMOS and PMOS.
\end{enumerate}

The active area of a circuit can be calculated by summing the contributions of all transistors, capacitors, and resistors. The formula for calculating the active area is as follows:
\begin{equation}
    Area = \sum_{i=1}^{n} (W_i \times L_i \times M_i \times P_i) + \sum_{j=1}^{c} (\text{C}_j \times 1089) + \sum_{k=1}^{r} (\text{R}_k \times 5)
\end{equation}
In this formula, \( W_i \) is the width of the \( i \)-th transistor, \( L_i \) is the length of the \( i \)-th transistor, \( M_i \) is the multiplicity of the \( i \)-th transistor, and \( P_i \) is the number of parallel transistors for the \( i \)-th transistor. \( \text{C}_j \) is the capacitance value of the \( j \)-th capacitor in picofarads (pF), and \( \text{R}_k \) is the resistance value of the \( k \)-th resistor in kilo-ohms (k$\Omega$). The total number of transistors, capacitors, and resistors is represented by \( n \), \( c \), and \( r \) respectively. 
 
This formula accounts for active area contributions from all transistors by multiplying their width, length, multiplicity, and parallel count, and includes contributions from capacitors and resistors, scaled by constants derived from real-world technology (1089 for pF and 5 for k$\Omega$) before taking the square root of the total sum.

\begin{table}[]
\caption{Text Representation of Testbench.}
\label{tab:Table of testbench}
\renewcommand{\arraystretch}{1.1}
\resizebox{\linewidth}{!}{
\begin{tabular}{c|l}
\hline
1 & .include ./netlist/NMCNR\_Pin\_3.txt                                                                         \\ \hline
2 & .include ./mosfet model/sky130\_pdk/libs.tech/ngspice/corners/tt.spice                                       \\ \hline
3 & .PARAM supply\_voltage = 1.8                                                                                 \\ \hline
4 & .temp 27                                                                                                     \\ \hline
5 & .PARAM PARAM\_CLOAD =10p                                                                                  \\ \hline
\multirow{7}{*}{6} & .PARAM    MOSFET\_10\_1\_L\_gm2\_PMOS=1 MOSFET\_10\_1\_M\_gm2\_PMOS=38  MOSFET\_10\_1\_W\_gm2\_PMOS=0.5 \\  
  & MOSFET\_23\_1\_L\_gm3\_NMOS=1   MOSFET\_23\_1\_M\_gm3\_NMOS=40    MOSFET\_23\_1\_W\_gm3\_NMOS=0.5          \\  
  & MOSFET\_8\_2\_L\_gm1\_PMOS=1   MOSFET\_8\_2\_M\_gm1\_PMOS=37    MOSFET\_8\_2\_W\_gm1\_PMOS=0.5             \\ 
  & MOSFET\_0\_8\_L\_BIASCM\_PMOS=1   MOSFET\_0\_8\_M\_BIASCM\_PMOS=40    MOSFET\_0\_8\_W\_BIASCM\_PMOS=0.5    \\ 
  & MOSFET\_17\_7\_L\_BIASCM\_NMOS=1   MOSFET\_17\_7\_M\_BIASCM\_NMOS=10    MOSFET\_17\_7\_W\_BIASCM\_NMOS=0.5 \\ 
  & MOSFET\_21\_2\_L\_LOAD2\_NMOS=1   MOSFET\_21\_2\_M\_LOAD2\_NMOS=25    MOSFET\_21\_2\_W\_LOAD2\_NMOS=0.411  \\ 
  & CAPACITOR\_0=63p CAPACITOR\_1=25p  CURRENT\_0\_BIAS=5u RESISTOR\_0=0.5k                                      \\ \hline
\end{tabular}
}
\vspace{-1em}
\end{table}

\section{The Analog Circuit Synthesis Problem} \label{sec:Problem}
AnalogGym aims to provide an open and practical testing suite for analog circuit synthesis.
In this section, we give an introduction to the analog circuit synthesis problem.

\subsection{Problem Definition}
Optimization of analog integrated circuits is a complex and multi-objective problem. The optimization process aims to adjust the circuit design parameters to improve performance, such as gain, bandwidth, power consumption, noise, and area. At the same time, the optimization must satisfy all constraints to ensure the reliability and stability of the circuit under various operating conditions.

The objective is to find the optimal vector of design parameters \(\mathbf{x}^*\) that ensures that the circuit meets all specifications under PVT variations by optimizing the design parameters \(\mathbf{x}\).
The mathematical formulation is as follows:
\begin{equation*}
\begin{aligned}
& \min_{\mathbf{x}} \quad {F}(\mathbf{x}, \mathbf{P, V, T}) = \left[ f_1(\mathbf{x}, \mathbf{P, V, T}), f_2(\mathbf{x}, \mathbf{P, V, T}), \ldots, f_k(\mathbf{x}, \mathbf{P, V, T}) \right] \\
& \text{s.t.} \quad g_i(\mathbf{x}, \mathbf{P, V, T}) \leq 0, \quad i = 1, \ldots, m \\
& \quad \quad \ h_j(\mathbf{x}, \mathbf{P, V, T}) = 0, \quad j = 1, \ldots, n \\
& \quad \quad \ (\mathbf{P, V, T}) \in \{(P_1, V_1, T_1), (P_2, V_2, T_2), \ldots, (P_N, V_N, T_N)\}
\end{aligned}
\end{equation*}

where \(\mathbf{x}\) is the vector of design parameters, including transistor dimensions, resistances, capacitances, bias currents, etc; \(\mathbf{F}(\mathbf{x}, \mathbf{P, V, T})\) is the multi-objective performance metrics vector; \(g_i(\mathbf{x}, \mathbf{P, V, T})\) are the inequality constraints; \(h_j(\mathbf{x}, \mathbf{P, V, T})\) are equality constraints; \((P_k, V_k, T_k)\) are the specific PVT combinations, \(k = 1, 2, \ldots, N\); 
\begin{table*}[]
\caption{Summary Table of Performance Metrics and Optimization Objectives for Multiple Analog Circuit.}
\label{tab:Table of Performance Metrics}
\renewcommand{\arraystretch}{1.2}
\resizebox{\textwidth}{!}{
\begin{tabular}{c|c|c|c|c|c|c|c|c|c}
\hline
\multicolumn{2}{c|}{Voltage   reference} &
  \multicolumn{2}{c|}{Sensing Front End} &
  \multicolumn{2}{c|}{Amplifier} &
  \multicolumn{2}{c|}{Low Dropout Regulator} &
  \multicolumn{2}{c}{Phase-Locked Loops} \\ \hline
\multirow{2}{*}{\begin{tabular}[c]{@{}c@{}}Voltage   \\      Difference (mV)\end{tabular}} &
  \multirow{2}{*}{$\downarrow$} &
  \multirow{2}{*}{\begin{tabular}[c]{@{}c@{}}Linearity \\      Relative Deviation (\%)\end{tabular}} &
  \multirow{2}{*}{$\downarrow$} &
  \textbf{Gain (dB)} &
  $\uparrow$ &
  Loop Bandwidth (MHz) &
  $\uparrow$ &
  \multirow{6}{*}{\begin{tabular}[c]{@{}c@{}}Loop \\      Bandwidth (Hz)\end{tabular}} &
  \multirow{6}{*}{$\uparrow$} \\ \cline{5-8}
 &
   &
   &
   &
  \textbf{Slew Rate (V/us)} &
  $\uparrow$ &
  \textbf{Current Efficiency (\%)} &
  $\uparrow$ &
   &
   \\ \cline{1-8}
\multirow{4}{*}{\begin{tabular}[c]{@{}c@{}}Line   \\      Sensitivity (\%/mV)\end{tabular}} &
  \multirow{4}{*}{$\downarrow$} &
  \multirow{3}{*}{\begin{tabular}[c]{@{}c@{}}Linearity   \\      Absolute Deviation ($^\circ$)\end{tabular}} &
  \multirow{3}{*}{$\downarrow$} &
  \multirow{2}{*}{\textbf{\begin{tabular}[c]{@{}c@{}}Gain-Bandwidth   \\      Product (MHz)\end{tabular}}} &
  \multirow{2}{*}{$\uparrow$} &
  \textbf{Dropout voltage (mV)} &
  $\downarrow$ &
   &
   \\ \cline{7-8}
 &
   &
   &
   &
   &
   &
  \textbf{Quiescent current (mA)} &
  $\downarrow$ &
   &
   \\ \cline{5-8}
 &
   &
   &
   &
  \begin{tabular}[c]{@{}c@{}}Input Offset \\ Voltage (mV)\end{tabular} &
  $\downarrow$ &
  \textbf{Line regulation (mV/V)} &
  $\downarrow$ &
   &
   \\ \cline{3-8}
 &
   &
  Noise ($Hz^{0.5}$) &
  $\downarrow$ &
  \textbf{Settling Time ($\mu$s)} &
  $\downarrow$ &
  \textbf{Recovery Time ($\mu$s)} &
  $\downarrow$ &
   &
   \\ \hline
\multirow{4}{*}{\textbf{\begin{tabular}[c]{@{}c@{}}Temperature   \\      Coefficient (ppm/$^{\circ}$C)\end{tabular}}} &
  \multirow{4}{*}{$\downarrow$} &
  \multirow{2}{*}{\textbf{Temperature Relative Inaccuracy (\%)}} &
  \multirow{2}{*}{$\downarrow$} &
  Integrated Noise (mV$_{rms}$) &
  $\downarrow$ &
  Integrated Noise (mV$_{rms}$) &
  $\downarrow$ &
  \multirow{4}{*}{\textbf{\begin{tabular}[c]{@{}c@{}}RMS Integrated \\      Jitter (ps)\end{tabular}}} &
  \multirow{4}{*}{$\downarrow$} \\ \cline{5-8}
 &
   &
   &
   &
  \multirow{2}{*}{\begin{tabular}[c]{@{}c@{}}Common-Mode   \\      Rejection Ratio (dB)\end{tabular}} &
  \multirow{2}{*}{$\downarrow$} &
  Output Capacitor (p) &
  $\downarrow$ &
   &
   \\ \cline{3-4} \cline{7-8}
 &
   &
  \multirow{2}{*}{\textbf{\begin{tabular}[c]{@{}c@{}}Temperature   \\      Absolute Inaccuracy ($^\circ$)\end{tabular}}} &
  \multirow{2}{*}{$\downarrow$} &
   &
   &
  \textbf{Voltage Deviation (mV)} &
  $\downarrow$ &
   &
   \\ \cline{5-8}
 &
   &
   &
   &
  \begin{tabular}[c]{@{}c@{}}Temperature   \\      Coefficient (ppm/$^{\circ}$C)\end{tabular} &
  $\downarrow$ &
  \textbf{Load Regulation ($\mu$V/mA)} &
  $\downarrow$ &
   &
   \\ \hline
\textbf{\begin{tabular}[c]{@{}c@{}}Power   \\      Consumption (mW)\end{tabular}} &
  $\downarrow$ &
  \textbf{\begin{tabular}[c]{@{}c@{}}Power \\      Consumption (mW)\end{tabular}} &
  $\downarrow$ &
  \textbf{\begin{tabular}[c]{@{}c@{}}Power   \\      Consumption (mW)\end{tabular}} &
  $\downarrow$ &
  \textbf{\begin{tabular}[c]{@{}c@{}}Power   \\      Consumption (mW)\end{tabular}} &
  $\downarrow$ &
  \textbf{\begin{tabular}[c]{@{}c@{}}Power   \\      Consumption (mW)\end{tabular}} &
  $\downarrow$ \\ \hline
\begin{tabular}[c]{@{}c@{}}Power   Supply \\      Rejection Ratio (dB)\end{tabular} &
  $\downarrow$ &
  \begin{tabular}[c]{@{}c@{}}Power Supply \\      Rejection Ratio (dB)\end{tabular} &
  $\downarrow$ &
  \begin{tabular}[c]{@{}c@{}}Power Supply \\      Rejection Ratio (dB)\end{tabular} &
  $\downarrow$ &
  \begin{tabular}[c]{@{}c@{}}Power Supply \\      Rejection (dB)\end{tabular} &
  $\downarrow$ &
  \begin{tabular}[c]{@{}c@{}}Power Supply \\      Rejection Ratio (dB)\end{tabular} &
  $\downarrow$ \\ \hline
Active Area ($\mu m^2$) &
  $\downarrow$ &
  Active Area ($\mu m^2$) &
  $\downarrow$ &
  Active Area ($\mu m^2$) &
  $\downarrow$ &
  Active Area ($\mu m^2$) &
  $\downarrow$ &
  Active Area ($\mu m^2$) &
  $\downarrow$ \\ \hline
\end{tabular}
}
\vspace{-1em}
\end{table*}
\begin{table}[]
\caption{Summary Table of Performance Constraints for Multiple Analog Circuit.}
\label{tab:Table of Constraints}
\renewcommand{\arraystretch}{1.2}
\resizebox{\linewidth}{!}{
\begin{tabular}{c|c|c|c|c}
\hline
\multicolumn{1}{c|}{Voltage   reference} &
  \multicolumn{1}{c|}{Sensing Front End} &
  \multicolumn{1}{c|}{Amplifier} &
  \multicolumn{1}{c|}{Low Dropout Regulator} &
  Phase-Locked Loops \\ \hline
\multirow{3}{*}{\begin{tabular}[c]{@{}c@{}}Reference \\      Voltage (mV)\end{tabular}} &
  \multicolumn{1}{c|}{\multirow{3}{*}{Calibration}} &
  \multicolumn{1}{c|}{\begin{tabular}[c]{@{}c@{}}Phase \\      Margin ($^{\circ}$)\end{tabular}} &
  \multicolumn{1}{c|}{\begin{tabular}[c]{@{}c@{}}Phase \\      Margin ($^{\circ}$)\end{tabular}} &
  \begin{tabular}[c]{@{}c@{}}Phase \\      Margin ($^{\circ}$)\end{tabular} \\ \cline{3-5} 
 &
  \multicolumn{1}{c|}{} &
  \multicolumn{1}{c|}{\begin{tabular}[c]{@{}c@{}}Input   Bias \\      Current (mA)\end{tabular}} &
  \multicolumn{1}{c|}{\begin{tabular}[c]{@{}c@{}}Output \\      voltage (V)\end{tabular}} &
  \begin{tabular}[c]{@{}c@{}}Output Frequency \\      Range (MHz)\end{tabular} \\ \cline{3-5} 
 &
  \multicolumn{1}{c|}{} &
  \multicolumn{1}{c|}{\begin{tabular}[c]{@{}c@{}}Load   \\      Capacitance (pF)\end{tabular}} &
  \multicolumn{1}{c|}{\begin{tabular}[c]{@{}c@{}}Output \\      Current (mA)\end{tabular}} &
  \begin{tabular}[c]{@{}c@{}}Reference \\      Frequency (MHz)\end{tabular} \\ \hline
\multicolumn{5}{c}{Supply Voltage (V)} \\ \hline
\multicolumn{5}{c}{Temperature Range ($^\circ$C)} \\ \hline
\end{tabular}
}
\vspace{-1em}
\end{table}
\subsection{PVT Variations}
\begin{figure}[htbp]
		\begin{center}
		\includegraphics[width=0.5\textwidth]{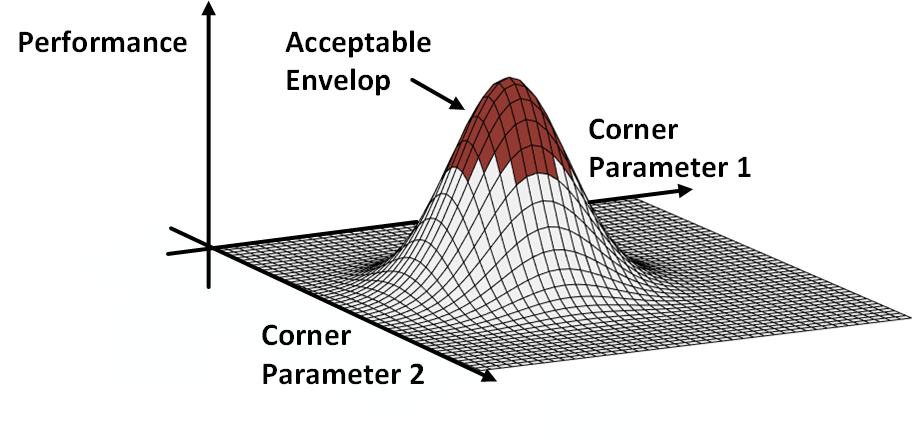}
		\end{center}
    \vspace{-1em} 
		\caption{The designer has to ensure that the performance of ICs is within a certain manageable range.}
		\label{fig:PVT_variations}
  \vspace{-1em} 
	\end{figure}

Assessing performance under PVT variations involves running simulations across all PVT corners with given design parameters \(\mathbf{x}\) to evaluate circuit performance.
Robust design requires managing performance variations within an acceptable range \cite{2024ESWA:li}, as shown in Figure \ref{fig:PVT_variations}, which includes the following types:
\begin{itemize}
\item[1)] Process corner modeling captures the variability in the parameters of N-MOSFET (NMOS) and P-MOSFET (PMOS) caused by variations in manufacturing processes, which also affect the electrical characteristics of transistors. The five commonly used process corners are slow NMOS+slow PMOS (SS), fast NMOS+fast PMOS (FF), slow NMOS+fast PMOS (SF), fast NMOS+slow PMOS (FS), and typical NMOS+typical PMOS (TT).
\item[2)] Voltage variations refer to fluctuations in power supply levels that integrated circuits encounter during operation. For example, in 130-nm process technology, typical voltage conditions include maximum voltage (1.32V), minimum voltage (1.08V), and nominal voltage (1.2V).
\item[3)] Temperature variations describe the influence of temperature changes on the transistor speed. 
These variations are evaluated at critical temperature points to ensure circuit reliability and functionality, which include high temperature (125$^\circ$C), low temperature (-40$^\circ$C), and room temperature (25$^\circ$C).
\end{itemize} 
 
\begin{table}[]
\caption{Summary Table of Physical Constraints under Different Technology Nodes.}
\label{tab:Table of Physical Constraints}
\renewcommand{\arraystretch}{1.2}
\resizebox{\linewidth}{!}{
\begin{tabular}{c|ccc}
\hline
\multicolumn{1}{c|}{Technology   Node}   & \multicolumn{1}{c|}{180nm}              & \multicolumn{1}{c|}{130nm}              & 22nm             \\ \hline
\multicolumn{1}{c|}{Length (nm)} & \multicolumn{1}{c|}{180 $ \sim $ 20000} & \multicolumn{1}{c|}{130 $ \sim $ 5000} & 22 $ \sim $ 500  \\ \hline
\multicolumn{1}{c|}{Width (nm)}  & \multicolumn{1}{c|}{220 $ \sim $ 50000} & \multicolumn{1}{c|}{200 $ \sim $ 10000} & 40 $ \sim $ 1000 \\ \hline
Multiplier                   & \multicolumn{3}{c}{1-1000 (Integer)}                                                                 \\ \hline
Input   Bias Current ($\mu$A)            & \multicolumn{3}{c}{0.5 $ \sim $ 100}                                                                 \\ \hline
\multicolumn{1}{c|}{Supply Voltage (V)} &
  \multicolumn{1}{c|}{1.62 $ \sim $ 1.98 (1.8 $\pm$ 10\%)} &
  \multicolumn{1}{c|}{1.08 $ \sim $ 1.32 (1.2 $\pm$ 10\%)} &
  \multicolumn{1}{c}{0.72 $ \sim $ 0.88 (0.8 $\pm$ 10\%)} \\ \hline
Temperature Range ($^\circ$C)            & \multicolumn{3}{c}{-25 $ \sim $ 125}                                                                 \\ \hline
Process Corner                           & \multicolumn{3}{c}{TT, FF, SS, FS, SF}                                                               \\ \hline
\end{tabular}
}
\vspace{-1em}
\end{table}
\subsection{Optimization Objectives and Constraints}
As shown in Table \ref{tab:Table of Performance Metrics}, a comprehensive table has been constructed to delineate the optimization objectives for multiple analog ICs. The table encompasses various performance metrics essential for evaluating and enhancing the efficacy of analog circuits. In the table, the symbol $\uparrow$ denotes that a larger value is preferable for the corresponding parameter, while the symbol $\downarrow$ indicates that a smaller value is optimal. In addition, the metrics highlighted in bold represent key optimization objectives.

The constraints include performance constraints, where the performance indicators of the circuit under each PVT condition must meet design requirements, and physical constraints, where the design parameters must be within reasonable physical limits. 
Table \ref{tab:Table of Constraints} presents the performance constraints for various analog circuits. Performance constraints serve as critical parameters that ensure circuits' proper and reliable operation within specified conditions and applications. Although these restrictions set essential boundaries, they can also be targeted for improvement. For example, optimizing the voltage reference to allow function over a wider temperature range can enhance the circuit's robustness.
Table \ref{tab:Table of Physical Constraints} presents the constraints on the dimensions of the transistor and the input bias current in different nodes of technology.
It is important to note that these constraints can vary slightly within the same technology node, depending on the specific PDK being used. 
Different PDKs may introduce variations in the allowable ranges for these parameters due to differences in manufacturing technologies and design optimizations.

\begin{table*}[]
\caption{Summary Table of Baseline Performances for Multiple Amplifiers under the Worst Case of PVT Variations.}
\label{tab:baseline}
\renewcommand{\arraystretch}{1.2}
\resizebox{\textwidth}{!}{
\begin{tabular}{c|c|c|c|c|c|c|c|c|c|c|c|c}
\hline
\begin{tabular}[c]{@{}c@{}}Topology\\ (180nm)\end{tabular} &
  \begin{tabular}[c]{@{}c@{}}PM\\ ($^\circ$)\end{tabular} &
  \begin{tabular}[c]{@{}c@{}}$C_{Load}$ \\ (pF)\end{tabular} &
  \begin{tabular}[c]{@{}c@{}}Gain\\ (dB)\end{tabular} &
  \begin{tabular}[c]{@{}c@{}}PSRR\\ (dB)\end{tabular} &
  \begin{tabular}[c]{@{}c@{}}CMRR\\ (dB)\end{tabular} &
  \begin{tabular}[c]{@{}c@{}}$v_n$\\ (mV$_{rms}$)\end{tabular} &
  \begin{tabular}[c]{@{}c@{}}vos\\ (mV)\end{tabular} &
  \begin{tabular}[c]{@{}c@{}}TC\\ (ppm/$^{\circ}$C)\end{tabular} &
  \begin{tabular}[c]{@{}c@{}}Ts\\ ($\mu$s)\end{tabular} &
  \begin{tabular}[c]{@{}c@{}}FOML\\ (V/$\mu$s*pF/mW)\end{tabular} &
  \begin{tabular}[c]{@{}c@{}}FOMS\\ (MHz*pF/mW)\end{tabular} &
  \begin{tabular}[c]{@{}c@{}}Active Area\\ (mm$^2$)\end{tabular} \\ \hline
Cascode\_Miller\cite{min_tan_cascode_2015}&
  \multirow{16}{*}{45 $ \sim $ 90} &
  10 &
  60 &
  \multirow{16}{*}{-60} &
  \multirow{16}{*}{-60} &
  \multirow{16}{*}{0.5} &
  \multirow{16}{*}{0.1} &
  \multirow{16}{*}{50} &
  \multirow{16}{*}{5} &
  30 &
  100 &
  100 \\ \cline{1-1} \cline{3-4} \cline{11-13} 
Alfio\_RAFFC \cite{alfio_dario_grasso_advances_2007}&
   &
  500 &
  \multirow{15}{*}{90} &
   &
   &
   &
   &
   &
  &
  1000 &
  1000 &
  200 \\ \cline{1-1} \cline{3-3} \cline{11-13} 
Fan\_SMC \cite{xiaohua_fan_single_2005}&
   &
   120 &
   &
   &
   &
   &
   &
   &
  &
  100 &
  500 &
  100 \\ \cline{1-1} \cline{3-3} \cline{11-13} 
HoiLee\_AFFC \cite{lee_active-feedback_2003}&
   &
   120 &
   &
   &
   &
   &
   &
   &
  &
  800 &
  500 &
  100 \\ \cline{1-1} \cline{3-3} \cline{11-13} 
Leung\_DFCFC1 \cite{2001:Leung} &
   &
   100 &
   &
   &
   &
   &
   &
   &
  &
  300 &
  200 &
  100 \\ \cline{1-1} \cline{3-3} \cline{11-13} 
Leung\_DFCFC2 \cite{2001:Leung} &
   &
   100 &
   &
   &
   &
   &
   &
   &
  &
  400 &
  1000 &
  100 \\ \cline{1-1} \cline{3-3}  \cline{11-13} 
Leung\_NMCF \cite{2001:Leung} &
   &
   100 &
   &
   &
   &
   &
   &
   &
  &
  60 &
  200 &
  200 \\ \cline{1-1} \cline{3-3} \cline{11-13} 
Leung\_NMCNR \cite{2001:Leung} &
   &
   100 &
   &
   &
   &
   &
   &
   &
  &
  20 &
  200 &
  500 \\ \cline{1-1} \cline{3-3} \cline{11-13} 
Peng\_ACBC \cite{peng_impedance_2011}&
   &
   500 &
   &
   &
   &
   &
   &
   &
  &
  100 &
  300 &
  100 \\ \cline{1-1} \cline{3-3} \cline{11-13} 
Peng\_IAC \cite{peng_ac_2004}&
   &
   150 &
   &
   &
   &
   &
   &
   &
  &
  500 &
  1500 &
  100 \\ \cline{1-1} \cline{3-3} \cline{11-13} 
Qu\_AZC \cite{qu_design-oriented_2017}&
   &
   18000&
   &
   &
   &
   &
   &
   &
   &
  5000 &
  60000 &
  100 \\ \cline{1-1} \cline{3-3} \cline{11-13} 
Qu\_LEC \cite{wanyuan_qu_173_2014}&
   &
   500 &
   &
   &
   &
   &
   &
   &
  &
  1000 &
  10000 &
  100 \\ \cline{1-1} \cline{3-3} \cline{11-13} 
Ramos\_PFC \cite{ramos_three_2002}&
   &
   130&
   &
   &
   &
   &
   &
   &
  &
  50 &
  100 &
  300 \\ \cline{1-1} \cline{3-3} \cline{11-13} 
Sau\_CFCC \cite{sau_siong_chong_cross_2012}&
   &
  500 &
   &
   &
   &
   &
   &
   &
  &
  1000 &
  5000 &
  100 \\ \cline{1-1} \cline{3-3} \cline{11-13} 
Song\_DACFC \cite{song_guo_dual_2011}&
   &
   800&
   &
   &
   &
   &
   &
   &
  &
  500 &
  1000 &
  100 \\ \cline{1-1} \cline{3-3} \cline{11-13} 
Yan\_AZ \cite{yan_nested-current-mirror_2015}&
   &
   15000&
   &
   &
   &
   &
   &
   &
  &
  500 &
  100000 &
  100 \\ \hline
\end{tabular}
}
\vspace{-1em}
\end{table*}
\subsection{Evaluation Metrics of Circuit}
In general, there are several widely used circuit figures of merits (FOMs) defined by human design expertise.
For example, the FOM for voltage references \cite{2020MEJ_liao,2023AEU_Yang} is used to evaluate various performance metrics comprehensively. 
This FOM covers all key optimization aspects of the voltage reference, including temperature coefficient (TC), line sensitivity (LS), power consumption, area, and voltage difference (\(\Delta V\)).
\begin{equation}
\label{FOM1}
\begin{split}
FOM(\boldsymbol{x}) = \frac{(T_{max} - T_{min})^2}{TC \times LS \times Power \times Area \times \Delta V}.
\end{split}
\end{equation}

However, many FOMs are insufficient for comprehensively characterizing circuit performance within algorithms. 
For example, commonly used FOMs for amplifiers, $\mathrm{FOM}_S$ and $\mathrm{FOM}_L$, are limited in scope. $\mathrm{FOM}_S$ measures the gain-bandwidth product (GBW) normalized by the load capacitance ($C_{Load}$) and power consumption, while $\mathrm{FOM}_L$ evaluates the slew rate (SR) normalized by the load capacitance and power consumption.
\begin{equation}
\begin{aligned}
\mathrm{FOM}_S &= \frac{\mathrm{GBW}[\text{MHz}] \cdot C_{\text{Load}}[\text{pF}]}{\text{Power}[\text{mW}]}, \\
\mathrm{FOM}_L &= \frac{\mathrm{SR}[\text{V}/\mu s] \cdot C_{\text{Load}}[\text{pF}]}{\text{Power}[\text{mW}]}.
\end{aligned}
\label{FOM}
\end{equation}

Although these FOMs are useful for evaluating specific aspects of circuit performance, they have significant limitations. For amplifiers, $\mathrm{FOM}_S$ and $\mathrm{FOM}_L$ do not fully encompass all critical performance metrics such as common-mode rejection ratio (CMRR), overall gain, and noise. Similarly, the FOM for LDOs might overlook important parameters such as the power supply rejection ratio (PSRR) and the transient response. 

In addition to individual performance and conventional FOM, AnalogGym also provides a reference metric,$FOM_{AMP}$,  for a comprehensive evaluation of FOM for amplifiers.
$FOM_{AMP}$ evaluates the optimization results by comparing the optimized and baseline performance metrics. The formula is as follows:
\footnotesize
\begin{equation}
\begin{aligned}
&\text{FOM}_{\text{AMP}} = \left( \frac{\text{PSRR}}{\text{PSRR}_{\text{ref}}} \cdot \frac{\text{CMRR}}{\text{CMRR}_{\text{ref}}} \cdot \frac{\text{Gain}}{\text{Gain}_{\text{ref}}} 
\cdot \frac{\text{FOMS}}{\text{FOMS}_{\text{ref}}} 
\cdot \frac{\text{FOML}}{\text{FOML}_{\text{ref}}} \right) \\ 
&\times \left( \frac{Ts}{Ts_{\text{ref}}} 
\cdot \frac{\text{Area}}{\text{Area}_{\text{ref}}} \right)^{-1} \\ 
&\times \left( 
\frac{v_n}{v_{n_{\text{ref}}}} \cdot \mathbb{I}(v_n > v_{n_{\text{ref}}}) 
\cdot \frac{TC}{TC_{\text{ref}}} 
\cdot \mathbb{I}(TC > TC_{\text{ref}})  
\cdot \frac{vos}{vos_{\text{ref}}} 
\cdot \mathbb{I}(vos > vos_{\text{ref}}) \right) ^{-1}
\end{aligned}
\end{equation}
\normalsize
where \(v_n\) is the integrated noise, \(Ts\) is the settling time, \(vos\) is the input offset voltage, \(lg\) refers to \(\log_{10}\), and \(\mathbb{I}(\cdot)\) is an indicator function, which equals 1 if the condition inside the parentheses is true, and 0 otherwise.

Table \ref{tab:baseline} illustrates the worst-case baseline performance of various AMPs under PVT variations, where the phase margin (PM) is a critical constraint that must be met to ensure the stability of the amplifier.
In particular, the Cascode\_Miller is a two-stage amplifier, while the others are three-stage amplifiers.
Furthermore, to maintain a fair comparison, the value of $C_{Load}$ should be consistent with the values specified in the table.
These baseline performances are derived from the analysis of the original papers proposing these topologies, identifying the fundamental criteria necessary for reliable circuit operation. 
The goal of algorithmic optimization is to produce better circuit designs where all performance metrics exceed their respective baseline performances.
Although $FOM_{AMP}$ is used to evaluate the final results of different optimization algorithms, it does not mean that the algorithms must use $FOM_{AMP}$ as their fitness function.
AnalogGym provides the interface to access the individual performance metrics and allows the circuit synthesizer to dynamically select its optimization objective.

\begin{table*}[]
\caption{Performance Summary of Single-Objective Optimization Across Different Topologies}
\label{tab:Performance_soo}
\renewcommand{\arraystretch}{1.2}
\resizebox{\textwidth}{!}{
\begin{tabular}{c|ccc|ccc|ccc|ccc}
\hline
Topology &
  \multicolumn{3}{c|}{NMCF} &
  \multicolumn{3}{c|}{NMCNR} &
  \multicolumn{3}{c|}{DFCFC1} &
  \multicolumn{3}{c}{DFCFC2} \\ \hline
Algorithm &
  \multicolumn{1}{c|}{GCNRL} &
  \multicolumn{1}{c|}{WEIBO} &
  cVTSBO &
  \multicolumn{1}{c|}{GCNRL} &
  \multicolumn{1}{c|}{WEIBO} &
  cVTSBO &
  \multicolumn{1}{c|}{GCNRL} &
  \multicolumn{1}{c|}{WEIBO} &
  cVTSBO &
  \multicolumn{1}{c|}{GCNRL} &
  \multicolumn{1}{c|}{WEIBO} &
  cVTSBO \\ \hline
PM   ($^{\circ}$) &
  \multicolumn{1}{c|}{62.2} &
  \multicolumn{1}{c|}{54.6} &
  59.5 &
  \multicolumn{1}{c|}{71.6} &
  \multicolumn{1}{c|}{63.2} &
  60.3 &
  \multicolumn{1}{c|}{65.2} &
  \multicolumn{1}{c|}{53.1} &
  59.4 &
  \multicolumn{1}{c|}{67.4} &
  \multicolumn{1}{c|}{59.7} &
  60.2 \\ \hline
Gain   (dB) &
  \multicolumn{1}{c|}{121.6} &
  \multicolumn{1}{c|}{126.7} &
  137.2 &
  \multicolumn{1}{c|}{119.4} &
  \multicolumn{1}{c|}{116.2} &
  123.4 &
  \multicolumn{1}{c|}{107.5} &
  \multicolumn{1}{c|}{124.3} &
  127.1 &
  \multicolumn{1}{c|}{112.6} &
  \multicolumn{1}{c|}{109.5} &
  128.5 \\ \hline
PSRR   (dB) &
  \multicolumn{1}{c|}{-72.3} &
  \multicolumn{1}{c|}{-71.5} &
  -70.3 &
  \multicolumn{1}{c|}{-68.2} &
  \multicolumn{1}{c|}{-65.2} &
  -58.3 &
  \multicolumn{1}{c|}{-72.4} &
  \multicolumn{1}{c|}{-59.5} &
  -66.2 &
  \multicolumn{1}{c|}{-52.4} &
  \multicolumn{1}{c|}{-67.4} &
  -54.8 \\ \hline
CMRR   (dB) &
  \multicolumn{1}{c|}{-72.4} &
  \multicolumn{1}{c|}{-74.8} &
  -69.4 &
  \multicolumn{1}{c|}{-74.6} &
  \multicolumn{1}{c|}{-51.6} &
  -46.2 &
  \multicolumn{1}{c|}{-63.7} &
  \multicolumn{1}{c|}{-68.2} &
  -67.5 &
  \multicolumn{1}{c|}{-57.3} &
  \multicolumn{1}{c|}{-60.3} &
  -53.6 \\ \hline
$v_n$   (mV$_{rms}$) &
  \multicolumn{1}{c|}{0.12} &
  \multicolumn{1}{c|}{0.08} &
  0.04 &
  \multicolumn{1}{c|}{0.5} &
  \multicolumn{1}{c|}{0.3} &
  0.6 &
  \multicolumn{1}{c|}{0.3} &
  \multicolumn{1}{c|}{0.5} &
  0.4 &
  \multicolumn{1}{c|}{0.9} &
  \multicolumn{1}{c|}{0.5} &
  0.4 \\ \hline
vos (mV) &
  \multicolumn{1}{c|}{0.05} &
  \multicolumn{1}{c|}{0.06} &
  0.03 &
  \multicolumn{1}{c|}{0.5} &
  \multicolumn{1}{c|}{0.4} &
  0.5 &
  \multicolumn{1}{c|}{0.4} &
  \multicolumn{1}{c|}{0.4} &
  0.2 &
  \multicolumn{1}{c|}{0.9} &
  \multicolumn{1}{c|}{0.6} &
  0.7 \\ \hline
TC  (ppm/$^{\circ}$C) &
  \multicolumn{1}{c|}{15.2} &
  \multicolumn{1}{c|}{10.4} &
  4.6 &
  \multicolumn{1}{c|}{44.5} &
  \multicolumn{1}{c|}{19.6} &
  27.2 &
  \multicolumn{1}{c|}{12.5} &
  \multicolumn{1}{c|}{10.6} &
  4.4 &
  \multicolumn{1}{c|}{37.2} &
  \multicolumn{1}{c|}{12.6} &
  10.4 \\ \hline
Ts   ($\mu$s) &
  \multicolumn{1}{c|}{9.7} &
  \multicolumn{1}{c|}{8.3} &
  3.6 &
  \multicolumn{1}{c|}{18.2} &
  \multicolumn{1}{c|}{12.5} &
  14.2 &
  \multicolumn{1}{c|}{19.6} &
  \multicolumn{1}{c|}{13.9} &
  16.4 &
  \multicolumn{1}{c|}{36.5} &
  \multicolumn{1}{c|}{39.4} &
  28.1 \\ \hline
FOML   (V/$\mu$s*pF/mW) &
  \multicolumn{1}{c|}{162.1} &
  \multicolumn{1}{c|}{152.7} &
  171.4 &
  \multicolumn{1}{c|}{114.3} &
  \multicolumn{1}{c|}{106.8} &
  175.4 &
  \multicolumn{1}{c|}{325.4} &
  \multicolumn{1}{c|}{312.6} &
  408.25 &
  \multicolumn{1}{c|}{275.4} &
  \multicolumn{1}{c|}{425.4} &
  771.9 \\ \hline
FOMS   (MHz*pF/mW) &
  \multicolumn{1}{c|}{406.4} &
  \multicolumn{1}{c|}{462.3} &
  485.9 &
  \multicolumn{1}{c|}{846.2} &
  \multicolumn{1}{c|}{782.4} &
  977.6 &
  \multicolumn{1}{c|}{754.2} &
  \multicolumn{1}{c|}{802.1} &
  767.23 &
  \multicolumn{1}{c|}{625.4} &
  \multicolumn{1}{c|}{294.6} &
  133.5 \\ \hline
Active   Area (mm$^2$) &
  \multicolumn{1}{c|}{142.5} &
  \multicolumn{1}{c|}{156.2} &
  139.7 &
  \multicolumn{1}{c|}{332.4} &
  \multicolumn{1}{c|}{302.4} &
  245.6 &
  \multicolumn{1}{c|}{104.5} &
  \multicolumn{1}{c|}{114.6} &
  84.1 &
  \multicolumn{1}{c|}{112.6} &
  \multicolumn{1}{c|}{104.5} &
  86.3 \\ \hline
FOM$_{AMP}$ &
  \multicolumn{1}{c|}{7.8} &
  \multicolumn{1}{c|}{9.5} &
  28.5 &
  \multicolumn{1}{c|}{2.5} &
  \multicolumn{1}{c|}{2.2} &
  4.2 &
  \multicolumn{1}{c|}{15.3} &
  \multicolumn{1}{c|}{20.4} &
  33.2 &
  \multicolumn{1}{c|}{3.6} &
  \multicolumn{1}{c|}{3.5} &
  4.1 \\ \hline
FEs   Consumption &
  \multicolumn{1}{c|}{5284.6} &
  \multicolumn{1}{c|}{826.1} &
  122.4 &
  \multicolumn{1}{c|}{6842} &
  \multicolumn{1}{c|}{796.2} &
  116.8 &
  \multicolumn{1}{c|}{6125.6} &
  \multicolumn{1}{c|}{892.4} &
  120.4 &
  \multicolumn{1}{c|}{5426.3} &
  \multicolumn{1}{c|}{916.4} &
  122.6 \\ \hline
Modeling   Time (min) &
  \multicolumn{1}{c|}{57.4} &
  \multicolumn{1}{c|}{274.6} &
  0.9 &
  \multicolumn{1}{c|}{78.5} &
  \multicolumn{1}{c|}{254.1} &
  0.8 &
  \multicolumn{1}{c|}{68.1} &
  \multicolumn{1}{c|}{316.5} &
  1.1 &
  \multicolumn{1}{c|}{61.4} &
  \multicolumn{1}{c|}{296.4} &
  1 \\ \hline
Simulation   Time (min)\textsuperscript{\textdagger} &
  \multicolumn{1}{c|}{16.8} &
  \multicolumn{1}{c|}{13.5} &
  5.2 &
  \multicolumn{1}{c|}{19.6} &
  \multicolumn{1}{c|}{15.9} &
  6.9 &
  \multicolumn{1}{c|}{19.4} &
  \multicolumn{1}{c|}{13.6} &
  7.2 &
  \multicolumn{1}{c|}{17.5} &
  \multicolumn{1}{c|}{13.7} &
  7.3 \\ \hline
Total   Runtime (min) &
  \multicolumn{1}{c|}{132.5} &
  \multicolumn{1}{c|}{341.6} &
  8.8 &
  \multicolumn{1}{c|}{158.4} &
  \multicolumn{1}{c|}{302.4} &
  8.8 &
  \multicolumn{1}{c|}{142.5} &
  \multicolumn{1}{c|}{395.2} &
  9.6 &
  \multicolumn{1}{c|}{139.6} &
  \multicolumn{1}{c|}{349.2} &
  9.5 \\ \hline
\end{tabular}
}
\begin{tablenotes}
\item\textsuperscript{\textdagger} \footnotesize Only the simulation time until algorithm convergence is considered.
\end{tablenotes}
\vspace{-1em}
\end{table*}
\begin{figure*}
    \centering
    \includegraphics[width=1\linewidth]{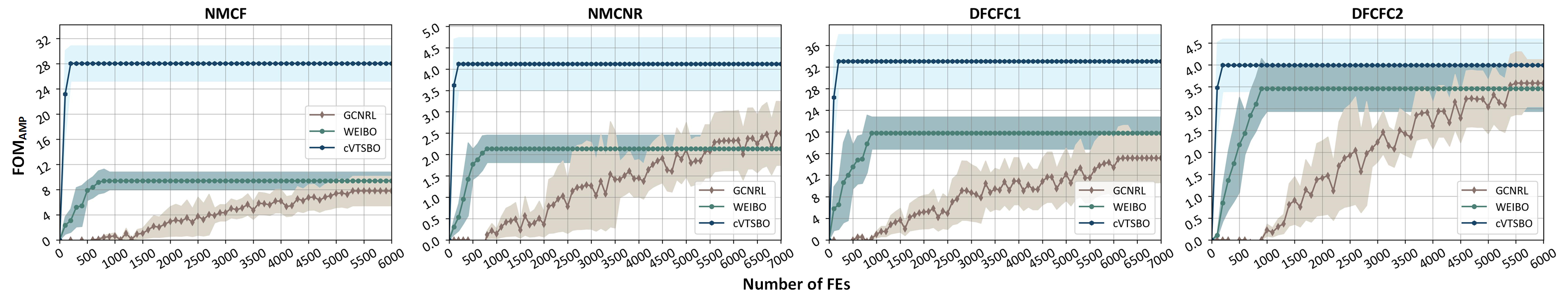}
    \vspace{-1em}
    \caption{Optimization curves of single-objective optimization algorithms under different topologies.}
    \label{fig:soo curves}
    \vspace{-1em}
\end{figure*}

\section{Algorithm Evaluation on AnalogGym} \label{sec:exp}
In this section, we optimize the AMPs within the testing suite based on the sky130 PDK of the 130nm technology node. 
All experiments are conducted on a Linux workstation with 2 NVIDIA GeForce RTX 4080 GPUs, 4 Intel Xeon Platinum 8260 CPUs, and 128 GB of memory. 
Notably, these test circuits are designed to meet the specifications required in practical design work. 
The circuit design may prioritize certain performance metrics, depending on the specific application. 

To validate the effectiveness and versatility of AnalogGym as a testing suite, we evaluate it using four representative circuit sizing algorithms, including reinforcement learning based on graph neural networks (GCNRL) \cite{2023ICCAD:GCN}, weighted expected improvement-based BO (WEIBO)~\cite{2018TCASI:lyu}, constrained Voronoi tree search BO (cVTS) \cite{2023DAC:cVTS_BO}, surrogate model-assisted evolutionary algorithms (SMEA) \cite{2021TCAD:ESSAB}, and multi-task evolutionary algorithms (MTEA) \cite{2024TCASI:Li}.
These algorithms were selected because they represent a wide range of optimization paradigms, including multi-objective and single-objective optimization, simulation-based methods, and surrogate-based approaches.
It is important to recognize that each algorithm has distinct focuses and strengths.
We evaluate these methods to verify the robustness and applicability of AnalogGym across various optimization strategies without fine-tuning the underlying algorithms.

\subsection{Experimental Setup}
Given the variability of results produced by different algorithms and the challenges in predicting optimization convergence, the experiment limited each algorithm to 1000 simulations. 

As a case study, we select four representative amplifiers: NMCNR, NMCF, DFCFC1, and DFCFC2. 
To accommodate the various paradigms of optimization algorithms, we design our experiments to include both single-objective and multi-objective optimization approaches. 
Each algorithm was run ten times to ensure fairness and reliability, and the average performance was calculated. 

The optimization objective for the single-objective optimization algorithms GCNRL and BO is to maximize \( FOM_{AMP} \).
To achieve a higher \( FOM_{AMP} \), we can impose stricter constraints on PSRR, CMRR, Ts, and Area. However, tightening these constraints will also increase the complexity of the optimization process.
Therefore, we define the problem as Equation \eqref{soo_problem}

For the multi-objective optimization algorithms SMEA and MTEA, which utilize hypervolume (HV) \cite{2006:HV} to measure the quality of solutions in the objective space, the objectives are defined as the Equation \ref{moo_problem} Note that PSRR and CMRR are represented as negative values, so the optimization effectively maximizes their absolute values.

\begin{flalign}
\label{soo_problem}
\quad \text{max} \quad &\text{FOML+FOMS} \notag &\\ 
\quad \text{s.t.} \quad & 45^\circ \leq PM \leq 90^\circ, 
 \text{Gain}  > 100 \, \text{dB}, 
 \text{Ts} < 1 \, \text{$\mu$s}, \notag &\\
& \text{PSRR}  < -60 \, \text{dB},
 \text{CMRR} < -60 \, \text{dB},  &\\
& v_n < 1 \, \text{mV}_{\text{rms}}, 
 v_{os} < 0.5 \, \text{mV}, \notag &\\
& \text{TC} < 100 \, \text{ppm}/^\circ \text{C}, 
 \text{Area}  < 500  \, \text{mm}^2, 
 C_{\text{Load}} = 100 \, \text{pF}. \notag &
\end{flalign}

\begin{equation}
\label{moo_problem}
\begin{split}
\text{max} \quad &\{\text{FOML}, \text{FOMS}, \text{Gain}, (\text{Area})^{-1}, (\text{Ts})^{-1}, \text{CMRR}, \text{PSRR}\}\\
\text{s.t.} \quad & 45^\circ \leq PM \leq 90^\circ, \\
& v_n < 1 \, \text{mV}_{\text{rms}}, 
 v_{os} < 0.5 \, \text{mV}, \\
& TC < 100 \, \text{ppm}/^\circ \text{C}, 
 C_{\text{Load}} = 100 \, \text{pF}.
\end{split}
\end{equation}

\begin{table*}[]
\caption{Performance Summary of Multi-Objective Optimization Across Different Topologies under PVT Variations}
\label{tab:Performance_moo}
\begin{tabular}{c|cc|cc|cc|cc}
\hline
Topology &
  \multicolumn{2}{c|}{NMCF} &
  \multicolumn{2}{c|}{NMCNR} &
  \multicolumn{2}{c|}{DFCFC1} &
  \multicolumn{2}{c}{DFCFC2} \\ \hline
Algorithm &
  \multicolumn{1}{c|}{SMEA} &
  MTEA &
  \multicolumn{1}{c|}{SMEA} &
  MTEA &
  \multicolumn{1}{c|}{SMEA} &
  MTEA &
  \multicolumn{1}{c|}{SMEA} &
  MTEA \\ \hline
PM   ($^{\circ}$) &
  \multicolumn{1}{c|}{57.4} &
  48.9 &
  \multicolumn{1}{c|}{61} &
  47.1 &
  \multicolumn{1}{c|}{49} &
  53.7 &
  \multicolumn{1}{c|}{59} &
  59.8 \\ \hline
Gain   (dB) &
  \multicolumn{1}{c|}{134.7} &
  126.0 &
  \multicolumn{1}{c|}{129.6} &
  132 &
  \multicolumn{1}{c|}{128.6} &
  104.5 &
  \multicolumn{1}{c|}{119.9} &
  107.5 \\ \hline
PSRR   (dB) &
  \multicolumn{1}{c|}{-90.5} &
  -101.4 &
  \multicolumn{1}{c|}{-91.2} &
  -95.6 &
  \multicolumn{1}{c|}{-74.5} &
  -72.6 &
  \multicolumn{1}{c|}{-82.7} &
  -75.6 \\ \hline
CMRR   (dB) &
  \multicolumn{1}{c|}{-102.1} &
  -82.5 &
  \multicolumn{1}{c|}{-87.5} &
  -73.7 &
  \multicolumn{1}{c|}{-83.2} &
  -71.5 &
  \multicolumn{1}{c|}{-90.4} &
  -74.5 \\ \hline
$v_n$   (mV$_{rms}$) &
  \multicolumn{1}{c|}{0.05} &
  0.05 &
  \multicolumn{1}{c|}{0.5} &
  0.4 &
  \multicolumn{1}{c|}{0.3} &
  0.1 &
  \multicolumn{1}{c|}{0.5} &
  0.2 \\ \hline
vos (mV) &
  \multicolumn{1}{c|}{0.03} &
  0.01 &
  \multicolumn{1}{c|}{0.6} &
  0.3 &
  \multicolumn{1}{c|}{0.5} &
  0.6 &
  \multicolumn{1}{c|}{0.4} &
  0.5 \\ \hline
TC  (ppm/$^{\circ}$C) &
  \multicolumn{1}{c|}{0.4} &
  0.2 &
  \multicolumn{1}{c|}{1.2} &
  0.9 &
  \multicolumn{1}{c|}{2.6} &
  2.7 &
  \multicolumn{1}{c|}{2.1} &
  2.0 \\ \hline
Ts   ($\mu$s) &
  \multicolumn{1}{c|}{0.8} &
  0.9 &
  \multicolumn{1}{c|}{3.4} &
  3.6 &
  \multicolumn{1}{c|}{0.9} &
  0.7 &
  \multicolumn{1}{c|}{1.4} &
  1.5 \\ \hline
FOML   (V/$\mu$s*pF/mW) &
  \multicolumn{1}{c|}{248.4} &
  195.6 &
  \multicolumn{1}{c|}{205.3} &
  176.2 &
  \multicolumn{1}{c|}{681.5} &
  647.5 &
  \multicolumn{1}{c|}{875.4} &
  797.4 \\ \hline
FOMS   (MHz*pF/mW) &
  \multicolumn{1}{c|}{584.6} &
  413.2 &
  \multicolumn{1}{c|}{815.5} &
  658.2 &
  \multicolumn{1}{c|}{1265.4} &
  832.5 &
  \multicolumn{1}{c|}{2368.1} &
  1925.6 \\ \hline
Active   Area (mm$^2$) &
  \multicolumn{1}{c|}{131.1} &
  147.6 &
  \multicolumn{1}{c|}{243.6} &
  279.6 &
  \multicolumn{1}{c|}{73.1} &
  72.6 &
  \multicolumn{1}{c|}{72.4} &
  70.5 \\ \hline
FOM$_{\text{AMP}}$ &
  \multicolumn{1}{c|}{443.2} &
  164.9 &
  \multicolumn{1}{c|}{53.8} &
  27.6 &
  \multicolumn{1}{c|}{2687.4} &
  1479.9 &
  \multicolumn{1}{c|}{4715.3} &
  2261.1 \\ \hline
FEs   Consumption &
  \multicolumn{1}{c|}{3156.3} &
  1000 &
  \multicolumn{1}{c|}{4265.7} &
  1000 &
  \multicolumn{1}{c|}{3967.2} &
  1000 &
  \multicolumn{1}{c|}{4257.1} &
  1000 \\ \hline
Modeling   Time (min) &
  \multicolumn{1}{c|}{10.4} &
  - &
  \multicolumn{1}{c|}{11.6} &
  - &
  \multicolumn{1}{c|}{16.5} &
  - &
  \multicolumn{1}{c|}{12.6} &
  - \\ \hline
Simulation   Time (min)* &
  \multicolumn{1}{c|}{4.4} &
  4.4 &
  \multicolumn{1}{c|}{5.3} &
  5.1 &
  \multicolumn{1}{c|}{6.8} &
  6.2 &
  \multicolumn{1}{c|}{5.4} &
  5.9 \\ \hline
Total   Runtime (min) &
  \multicolumn{1}{c|}{27.6} &
  20.4 &
  \multicolumn{1}{c|}{35.8} &
  22.8 &
  \multicolumn{1}{c|}{34.1} &
  19.5 &
  \multicolumn{1}{c|}{32.5} &
  20.4 \\ \hline
\end{tabular}
\begin{tablenotes}
\item *\footnotesize The simulations were conducted using 64-core parallel processing.
\end{tablenotes}
\vspace{-1em}
\end{table*}
\begin{figure*}
    \centering
    \includegraphics[width=1\textwidth]{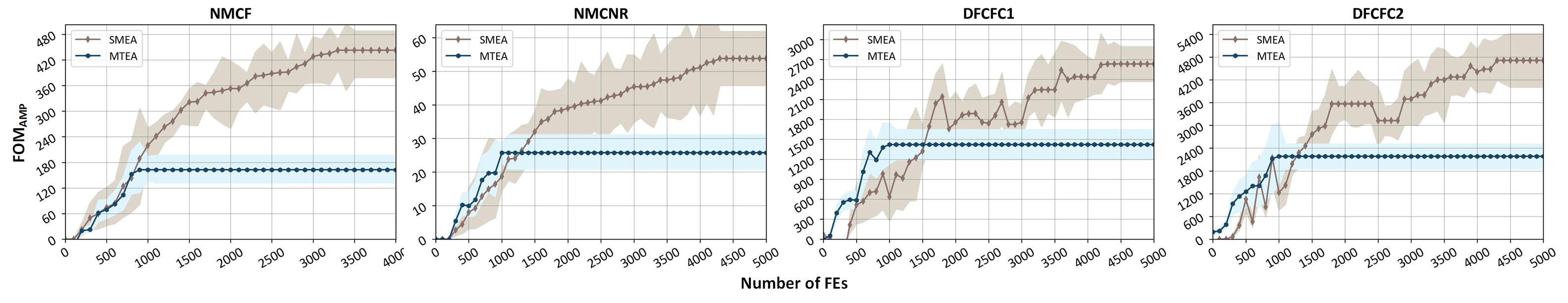}
    \vspace{-1em}
    \caption{Optimization curves of multi-objective optimization algorithms under different topologies.}
    \label{fig:moo curves}
    \vspace{-1em}
\end{figure*}

The input parameters include channel width (W), length (L), and multiplier (M), as well as capacitance (C), resistance (R), and bias current (I\(_{\text{bias}}\)). In the sky130 PDK, the specific ranges for these parameters are as follows:
\begin{equation}
\begin{split}
&W \in [0.2, 10] \, \mu\text{m}, \quad L \in [0.13, 1] \, \mu\text{m}, \quad M \in [1, 100], \\
&C \in [1, 100] \, \text{pF}, \quad R \in [0.1, 1000] \, \text{K}\Omega, \quad I_{\text{bias}} \in [1, 40] \, \mu\text{A}
\end{split}
\end{equation}

To meet the area constraint, we choose smaller parameter ranges. 
For larger input ranges, refer to Table \ref{tab:Table of Physical Constraints}.

Original SMEA and MTEA algorithms target optimization with consideration of PVT variations. 
We considered PVT variations using a worst-case-driven approach, ensuring the circuits met performance requirements under the most challenging specified conditions.
We designed the experiment to include four specific PVT corners, as follows:
\begin{equation}
\begin{split}
& (\mathbf{P, V, T}) \in \{ (SS, 1.08V, -25^\circ C), (FF, 1.32V, 125^\circ C),  \\
& \quad (SF, 1.32V, -25^\circ C), (FS, 1.08V, 125)\}.
\end{split}
\end{equation}

\subsection{Single-Objective Optimization}
Table \ref{tab:Performance_soo} and Figure \ref{fig:soo curves} present the performance of circuits optimized using BO and GCNRL, also showcasing their efficiency by showing the consumption of fitness evaluation (FE), the simulation time, the modeling time and the total runtime. 
FE consumption is recorded until the algorithm reaches convergence. 
For RL, FEs encompass the number of interactions between the agent and the environment, meaning each time an action is taken and feedback is received. 
For BO, FEs refer to the number of actual objective function evaluations performed at the recommended points

\subsection{Multi-Objective Optimization}
 Table \ref{tab:Performance_moo} and Figure \ref{fig:moo curves} shows the optimized circuit performance with two EAs.
EAs leverage parallelism by evaluating multiple candidate solutions (individuals) concurrently within a population, leading to parallel simulations. 
For EAs, FEs are calculated as the number of individuals in the population multiplied by the number of generations, along with additional evaluations needed to update and validate the surrogate model. 
MTEA does not achieve convergence within 1000 simulations in four topologies, resulting in an FE consumption of 1000 for each experiment.

\section{Discussion and Future Directions} \label{sec:discussion}

The difficulty in getting started with analog circuit optimization is mainly due to the lack of open-source netlists, simulators, and PDKs, combined with the required specialized circuit knowledge.
In response to these challenges, we have proposed a testing suite to make analog circuit optimization more accessible. The essence of AnalogGym is to transform the problem into a more open-source, mathematically describable optimization task. However, this transformation is far from straightforward.

\textbf{Applying different algorithms:}
We identify several points in evaluating different optimization methods.
1) Applying strict performance constraints to achieve a higher \( FOM_{AMP} \) can trap the optimization algorithm in local optima and restrict the search space, leading to longer convergence times and suboptimal designs.
2) \( FOM_{AMP} \) as the optimization objective can cause convergence issues, as it involves normalizing and multiplying performance metrics, increasing sensitivity to variations, and introducing noise that complicates the optimization process.
3) Prioritizing time constraints in the optimization process is crucial. Transient instability, such as oscillations or prolonged settling, can extend simulation times and reduce efficiency, leading to inaccurate evaluation of other performance metrics.

\begin{figure}
    \centering
    \includegraphics[width=1\linewidth]{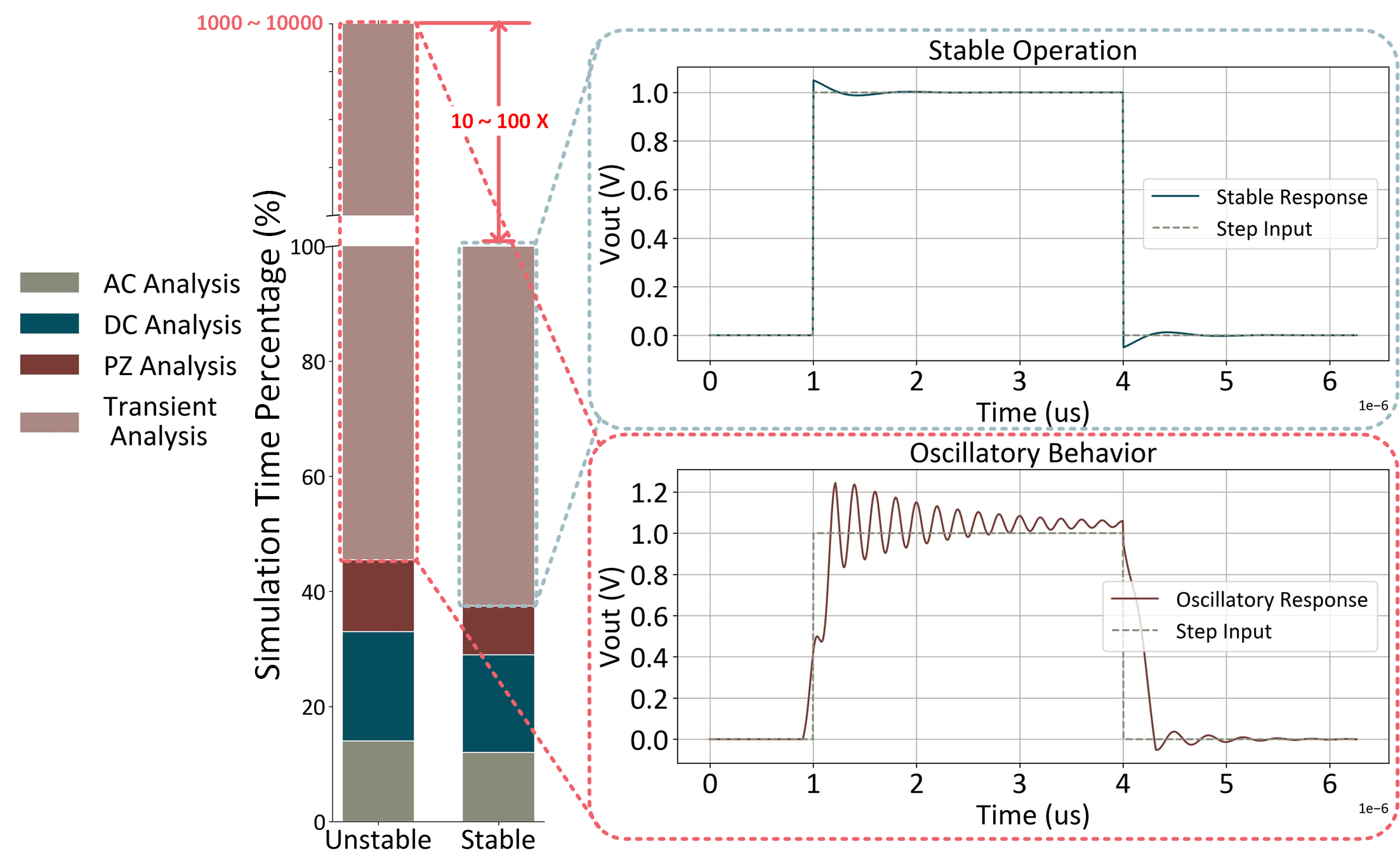}
    \caption{Comparison of Simulation Time for AMP Under Stable and Unstable Transient Conditions}
    \label{fig:unstable_transient}
\end{figure}

\textbf{Multi-objective optimization:}
One of the most challenging aspects of analog circuit optimization is balancing multiple objectives simultaneously, such as gain, settling time, and area, while satisfying a variety of constraints. 
SOO combines multiple objectives, but tuning weights and penalties is challenging.
Managing the Pareto front in MOO is also challenging due to the exponential computational complexity, particularly when calculating HV. 
Although \( FOM_{AMP} \) provides a useful metric for comparing circuit performance, these challenges still make it difficult to compare and evaluate different algorithms under complex conditions consistently.

When using multi-objective optimization (MOO) algorithms on AnalogGym, the increased computational complexity from multiple objectives must be considered. The HV metric, which measures the volume of the objective space dominated by a solution set, significantly increases complexity, with time complexity escalating from \(O(n \log n)\) in two dimensions to \(O(n^m)\) in \(m\) dimensions, extending runtime. Additionally, many objectives can introduce optimization bias, where certain objectives are prioritized, leading to suboptimal trade-offs and less balanced solutions.

\textbf{Transient stability:}
Optimizing dynamic performance in multi-stage AMPs is challenging due to difficulties in maintaining transient stability. Complex interactions between stages can cause phase shifts and feedback loops, leading to instability. Consequently, metrics like settling time and slew rate, crucial for dynamic performance, become difficult to model and optimize during oscillations due to the circuit's non-linear and time-varying behavior.

This sensitivity complicates metric quantification, as values may not stabilize, providing inconsistent feedback that hinders optimization. Additionally, extended simulation times needed to resolve oscillations increase complexity, with optimization potentially slowing by 10 to 100 times due to higher computational costs, as shown in Figure \ref{fig:unstable_transient}.

\subsection{Future Directions}

Future development of AnalogGym should aim to express circuit optimization challenges.
\begin{enumerate}
    \item Expanding AnalogGym to include more comprehensive evaluation and comparison methods, covering a broader range of circuit types and algorithmic paradigms. 
    \item Introducing layout considerations. Layout design is also crucial to circuit performance. In the future, we plan to augment AnalogGym with layout generation capability utilizing frameworks such as MAGICAL~\cite{2019ICCAD:Xu}.
    \item Adding support for topology search, which precedes device sizing, is planned for AnalogGym. Future developments will include an interface allowing users to modify circuit topologies, thereby expanding the search space.
\end{enumerate}
\section{Conclusion} \label{sec:conclusion}

In this paper, we propose AnalogGym, addressing the critical need for a standardized evaluation framework in analog circuit synthesis. By providing a comprehensive and open-source testing suite, it facilitates fair comparisons, enhances reproducibility, and bridges the gap between academic research and industrial applications. 

\bibliographystyle{unsrt}

\bibliography{ref}

\end{document}